\documentclass[12pts]{article}
\usepackage[displaymath]{lineno}
\usepackage[utf8]{inputenc}
\usepackage[usenames]{color}
\usepackage{graphicx}
\usepackage{amsmath}
\usepackage{commath}
\usepackage[version=3]{mhchem}
\usepackage[varg]{txfonts}
\usepackage{tcolorbox}
\usepackage{siunitx}
\usepackage{xspace}
\usepackage{natbib}
\usepackage{filecontents}

\begin{document}
\title{On the electrostatic field created at ground level by a halo}

\author{
  F. J. P\'erez-Invern\'on$^{1}$,
  F. J. Gordillo-V\'azquez$^{1}$,
  A. Luque$^{1}$. \\
\textit{$^{1}$Instituto de Astrof\'isica de Andaluc\'ia (IAA),} \\
   \textit{CSIC, PO Box 3004, 18080 Granada, Spain.}
\footnote{Correspondence to: fjpi@iaa.es. 
Article published in Geophysical Research Letters. }
}
\date{}
\maketitle

\begin{abstract}
  We investigate the effect of halo activity on the electrostatic 
field measured at ground level. We use electrostatic arguments 
as well as self-consistent simulations
to show that, due to the screening charge in the ionosphere,
the distant electrostatic field created by the uncompensated charge 
in a thundercloud decays exponentially rather than as the third power 
of the distance.  Furthermore,
significative ionization around the lower edge of the ionosphere 
slightly reduces the electrostatic field at ground level.
We conclude that halos do not extend the range of detectability 
of lightning-induced electrostatic fields. 

\end{abstract}

%\begin{article}
\section{Introduction}

\label{sect:intro}
Remotely detecting lightning strokes is essential to minimize the risks 
involved in electrical storms and other types of severe weather.  Most of the 
lightning detection systems currently in operation 
\citep{Cummins1998/JGR, Dowden2002/JASTP, Betz2009/AtmRe}
rely on the measurement of the radiation field emitted by the rapidly varying 
current pulse of the lightning discharge 
\citep{Cummins2009/IEEETransEC, Rakov2013/SGeo}.  
On the other hand, the electrostatic field created by a net charge 
within the thundercloud 
is rarely employed for lightning detection: whereas the radiation field decays with the inverse of the distance to the source, the electrostatic field decays as the inverse of the third power of distance for intermediate distances and much faster at longer distances.  Due to this faster decay it is impractical in most situations to measure the electrostatic fields at distances longer than about \SI{100}{km}.

Recently \cite{Bennett2013/PhRvL} reported the detection of 
lightning-produced electrostatic fields at distances of up to about 
\SI{300}{km} and thus a possible violation of the cubic decay law.  
\cite{Bennett2014/JASTP} explained this observation as resulting from the 
extended disk of charge induced by the thundercloud charge in the 
lower boundary of the ionosphere.  In their model this disk is 
associated to a halo: a well-studied diffuse light emission
closely below the ionosphere created also by lightning quasi-electrostatic  fields \citep{Barrington-Leigh2001/JGR,Pasko2010/JGRA}.  As the 
horizontal extension of halos reaches about one hundred kilometers, it is 
reasonable to claim, as \cite{Bennett2014/JASTP}, that they extend the reach of 
electrostatic fields at the ground.  Furthermore, because the ground-level 
electric field 
created by the charge in the ionosphere has the opposite polarity 
to the field created by the cloud charge, one can also argue that 
the electrostatic field due to a lightning discharge reverses its polarity
as it is measured at increasing distance from the source.  This
reversal was also reported by \cite{Bennett2013/PhRvL}.

These observations and models motivated us to investigate in greater detail 
the effect of halos on the electric field measured at the ground.  A key 
element that was missing in the above explanations is that the charge 
induced at the bottom of the ionosphere is a self-consistent response to 
the electrostatic field created by the cloud charge.  In other words, it is
a screening charge that reduces the electric field in the conducting ionosphere.
As we describe below, we found that a screening charge at the boundary of the ionosphere does not extend the range of its causative electrostatic field; rather, it always reduces the magnitude of this field.  Furthermore, the orientation of the field cannot be reversed due to the presence of this screening charge.

To reach this conclusion we first review the physics of halos 
and discuss how the
upper-atmospheric electrical activity may influence the field at the surface.  Then we present electrostatic arguments of why a screening charge at the ionosphere does not enhance the distant electric field.  These arguments are then applied to our main results, where we use a self-consistent, quasi-electrostatic model of the response of the ionosphere to a lightning discharge.  Within a wide range of causative charge-moments, we consistently find that the charge accumulated on the ionosphere reduces the distant field at ground level relative to the raw dipolar field created by the charges in the thurdercloud.  We therefore conclude that halos are not responsible of the field enhancements observed by \cite{Bennett2013/PhRvL}.

\section{The physics of halos}
\label{sect:halos}
Halos are a type of transient luminous events (TLEs) in the upper 
atmosphere, a family of light-emitting phenomena associated with lightning that 
were first described by \cite{Franz1990/Sci} and that besides halos includes 
sprites, elves, blue jets and giant blue jets 
\citep{Ebert2010/JGRA,Pasko2012/SSRv,Liu2015/JASTP}.  TLEs in the upper atmosphere (halos, sprites and elves) owe their existence 
to the rarefied air density at high altitude: as electrons experience fewer collisions with air molecules, they are more readily accelerated to high energies, and are thus capable of ionizing molecules or exciting them into light-emitting states.  Since the lower daytime ionosphere prevents the penetration of electric fields to high altitudes, TLEs exist mostly during nighttime.  In any case, the observation of daytime TLEs would be problematic because their emissions would be swamped by scattered sunlight.

Halos are one of the most frequent types of TLEs: they are diffuse, saucer-shaped light emissions at \SIrange{80}{90}{km} of altitude with diameters of about \SI{100}{km} that propagate downwards and last about \SI{1}{ms}.  In a halo electrons obviously reach energies high enough to excite substantial numbers of molecules into radiating states, namely into $\ce{N2}(B^3\Pi_g)$, which radiates in the first positive band of nitrogen.  It is however not so clear whether they also have enough energy to cause substantial ionization.  This is certainly the case when the halo initiates a sprite, as was studied by \cite{Luque2009/NatGe}, 
\cite{Qin2014/NatCo} and \cite{Liu2015/NatCo/1}.  Besides, \cite{Kuo2013/JGRA} 
detected signatures of ionization in one halo not associated with a sprite.  We therefore conclude that, although there may be some visible halos without a substantial effect on the upper-atmospheric electron density, many others do increase this density and thus the electrical conductivity below the lower edge of the ionosphere.  The increase of conductivity is not necessarily simultaneous to the luminosity of the halo: as investigated by \cite{Luque2012/NatGe}, \cite{Liu2012/JGRA} and \cite{Parra-Rojas2013/JGRA}, delayed electron detachment causes conductivity enhancements on timescales of \SIrange{10}{100}{ms}, long after the luminosity has decayed.

The increase of conductivity caused by an ionizing halo can also be viewed as a transient and localized lowering of the ionosphere, which according to e.g. \cite{Luque2012/NatGe} and \cite{Liu2012/JGRA}, can reach as low as \SI{70}{km} of altitude.  It is this descent of the ionosphere's edge that may plausibly lead to an extended horizontal range of the electrostatic field created by a lightning stroke.  Note that the ionosphere is present regardless of any halo activity and therefore there is always some screening charge in response to an electrostatic field: it is the extension and magnitude of this charge that may be affected by the presence of a halo.

\section{Electrostatics of a halo}
\label{sect:electro}
Let us now consider how a lower ionosphere influences the electrostatic field at ground level.  It is useful to first analyze a simplified system where the charge that the stroke leaves in and around the thundercloud can be modeled as a point charge sitting between two perfectly conducting surfaces representing the ground and the lower edge of the ionosphere.  Although the voltage difference between ground and ionosphere is about \SI{+250}{kV} \citep{Rycroft2000/JASTP} we assume that our two conductors are at the same potential.  This choice is mainly justified by the roughly exponential increase of the conductivity in the atmosphere for increasing altitude.  This exponential profile causes the potential drop to be almost completely concentrated at low altitude so changes around the ionosphere have a negligible effect on the electric field caused by this potential bias.  In addition, in the observations by \cite{Bennett2014/JASTP} the DC bias was filtered out by a \SI{1}{Hz} high-pass filter.

We therefore consider a point charge placed between two conducting, grounded electrodes.  In the simplest geometry of this setup both conducting surfaces are plane and parallel.  In that case the electric field  can be calculated by summing an infinite series of image charges.  For the vertical component of the electric field at ground level at a plane distance $r$ from the thundercloud we find
\begin{linenomath*}
\begin{equation}
  E_z(r) = -\frac{Q}{2\pi\epsilon_0}\sum_{n=-\infty}^{\infty} 
     \frac{h + 2nL}{\left[(h + 2nL)^2 + r^2\right]^{3/2}},
     \label{images}
\end{equation}
\end{linenomath*}
where $Q$ is the total charge in the thundercloud, located at an altitude $h$ above ground, $L$ is the ground-ionosphere separation and $\epsilon_0$ is the vacuum permittivity.

\begin{figure}
\includegraphics[width=0.8\columnwidth]{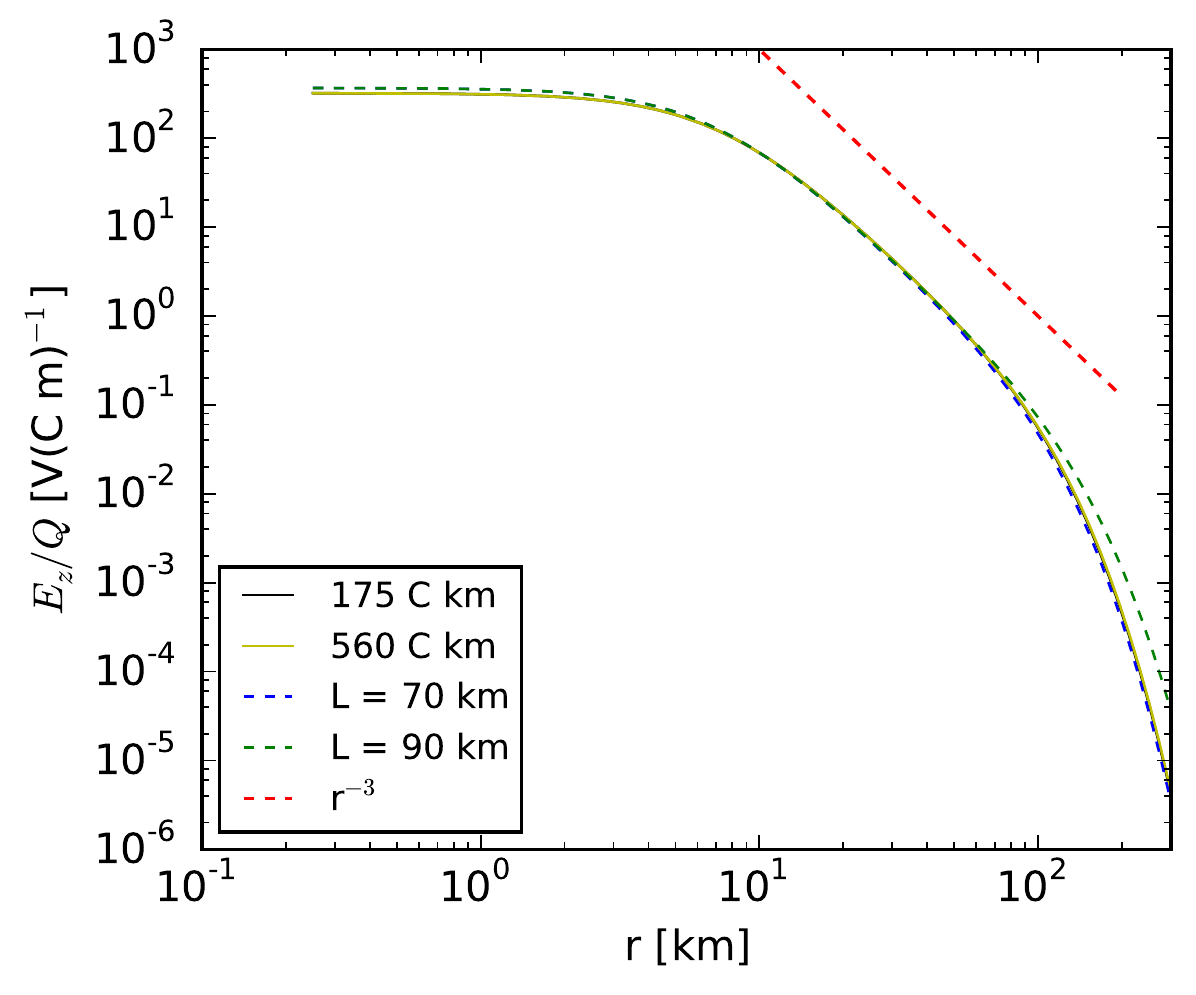}

\caption{\label{fig:ground-field}
  Vertical component of the electric field at ground level divided by its causative charge $Q$.  We show two evaluations of (\ref{images}) (dashed lines) where the ionosphere is represented by a planar, perfect conductor either at $L=\SI{90}{km}$ or $L=\SI{70}{km}$ as well as the outcome of two simulations described in section \ref{sect:simulations} (coincident solid lines).  For the simulations, we plot the electric field 10 ms after the start of the discharge.
We also provide an arbitrarily placed reference line to illustrate the slope of a $r^{-3}$ decay.
}
\end{figure}

We evaluate (\ref{images}) numerically by truncating the infinite series.  As the sum converges very slowly for large $r$, we extended the sum to all terms with
$-\num{e5} < n < \num{e5}$.  In figure~\ref{fig:ground-field} we see that the ground-level electric field $E_z(r)$ goes through three regimes as the distance to the causative lightning $r$ increases.  When $r \approx h$ the field is approximately constant; we are not interested in this range, where our simplification of the charge distribution in the cloud as a single point charge breaks down.  For intermediate distances where $r \approx L \gg h$ the field decays as $r^{-3}$ because (\ref{images}) is dominated by the $n=0$ term; in this range the field is approximately dipolar.  Finally we see that for $r \gg L$ the decay of the field is exponential and hence much faster than the dipolar field.  This is a key observation, since it shows that the presence of the upper electrode induces a much faster decay of the  distant field. Furthermore, we also found that as $L$ decreases,  $|E_z(r)|$ also decreases for all $r$, as long as $L > h$.  In our context, this means that a lower ionosphere implies a lower electrostatic field at ground level.

The exponential decay of the series (\ref{images}) for large $r$ can be proven
analytically by using techniques originally developed to calculate the electric field created by a crystal lattice of ions with alternating charges (see e.g. \cite{Borwein2013/book}, p. 5ff).  This derivation fall out of the scope of this letter but we can summarize it as follows: the Poisson summation formula transforms (\ref{images}) into a series that converges much faster and can be truncated to a single term for $r \to \infty$; asymptotically, this term decays exponentially.

\begin{figure}
\includegraphics[width=0.7\columnwidth]{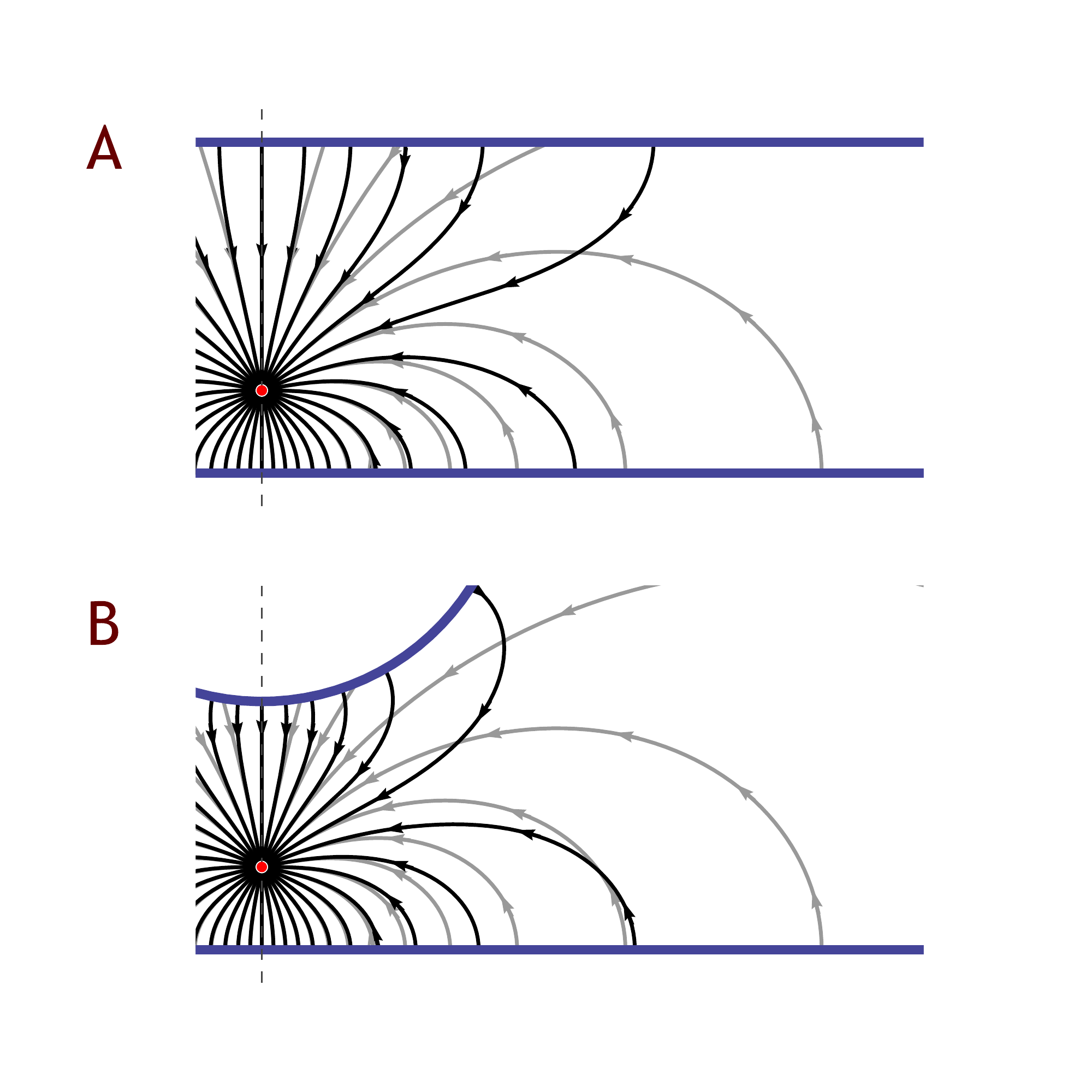}
\caption{\label{fig:fieldlines}
Field lines created by a negative point charge between two perfect 
conductors. In (A) the field lines created by two planar, infinite are plotted as black lines.  For reference we also plot the field lines created when the upper conductors in infinitely removed from the charge.  We see that the upper conductor, by ``attracting'' field lines reduces the line density in the lower conductor.  In terms of our problem, this means that a lower ionosphere reduces the electric field at the surface.  In (B) we consider a curved upper electrode, as would be created by a realistic halo.  Here also the black lines indicate the field lines in this geometry and the gray lines provide a reference where the upper electrode is absent.  The field is enhanced directly below the vertex but away from it the field in the lower conductor decreases. 
}
\end{figure}

The exponential decay of the electric field for long distances allows us to estimate the total charge in the ionosphere, $Q_\text{i}$ as follows.  For $r\gg L$ we can view the system as the sum of two dipoles: the dipole $2Qh$ created by the causative charge and its image on the ground and the dipole $2Q_\text{i}L$ created by the charge in the halo and its image.  Since at long distances there is no dipolar field, both contributions cancel:
\begin{linenomath*}
\begin{equation}
  Q_\text{i} = -\frac{Q h}{L}.
  \label{Qhalo}
\end{equation}
\end{linenomath*}

For our two-electrode model we can visualize the reduction of the electric field in the lower electrode caused by the upper electrode by looking at electric field lines, as shown in figure~\ref{fig:fieldlines}A.  The boundary conditions
force these to be perpendicular to both conducting surfaces and therefore they curve outwards.  The upper conductor ``attracts'' the field lines at the expense of the line density at the lower conductor, yielding a lower field there.

Another point that can be illuminated by looking at the electric field lines is whether the screening charge in the ionosphere can reverse the orientation of the electric field at ground level.  For concreteness, assume that the net charge in the thundercloud is negative.  Now suppose that the electric field at some point in the lower surface points downward, thus marking the endpoint of a field line.  The startpoint of this line cannot be the space charge, which is negative, nor the upper electrode, since that would imply a potential difference between the electrodes.  Finally, using arguments similar to those used above, one can show that in this configuration the radial electric field also decays exponentially and therefore the field line cannot extend indefinitely outwards.  We conclude that it is impossible for the field to point downwards at the lower electrode.  This means that the charge induced in the ionosphere cannot be high enough to reverse the polarity of the electric field at ground level.

There are, however, two aspects where the model of parallel conductors oversimplifies the physics of an actual halo: 
\begin{enumerate}
  \item Rather than an infinitely sharp, perfect conductor, the ionosphere consists of a smoothly increasing electric conductivity.
  \item After a lightning stroke the ionosphere does not descend uniformly: the region directly above the stroke is ionized more intensely and a bulge emerges from the lower edge of the ionosphere.
\end{enumerate}
In the next section we will describe self-consistent simulations where these two simplifications are removed. Nevertheless, it is worth discussing qualitatively the reasons why they do not invalidate our previous argument.

\begin{enumerate}
  \item The first issue can be quickly dismissed: a finite conductivity slows down the transport of charge to the lower boundary of the ionosphere.  But we have seen that the effect of this screening charge is to decrease the electric field at ground level so a slower charge accumulation merely implies that this decrease is weaker and slower.
  \item The second issue is potentially more problematic since the curvature of the ionosphere's edge enhances the electric field directly below the point of highest curvature, which is vertically aligned with the causative lightning (see figure~\ref{fig:fieldlines}B).  However, we are interested in the electric field at locations farther than about \SI{100}{km} from the parent lightning.  Since this is also roughly the horizontal span of a halo, we expect that the curvature effect at those distances is negligible or even reversed, actually weakening the electric field.  However, we cannot exclude that the distant electric field is enhanced by other sources of ionization away from the causating lightning, such as inhomogeneities caused by gravity waves \citep{Liu2015/NatCo} or the electromagnetic pulse (EMP) emitted by the lightning stroke.
\end{enumerate}
  
\section{Self-consistent simulations}
\label{sect:simulations}
Let us now flesh out the above arguments with realistic, self-consistent simulations.  We use a cylindrically symmetrical density model for electron transport in the mesosphere and lower ionosphere, similar to previous models by e.g. \cite{Luque2009/NatGe,Neubert2011/JGRA,Liu2015/NatCo} and \cite{Qin2014/NatCo} (this type of models were reviewed by e.g. \cite{Pasko2010/JGRA} and \cite{Luque2012/JCoPh}).  In our model, electrons drift in a self-consistent electric field and interact with neutrals within a minimal chemical scheme that includes impact ionization, dissociative attachment and associative detachment:
\begin{linenomath*}
\begin{subequations}
 \label{reactions}
\begin{align}
   \cee{e + N2 & -> N2+ + 2e}, \label{n2ion} \\ 
   \cee{e + O2 & -> O2+ + 2e}, \label{o2ion} \\ 
   \cee{e + O2 & -> O + O-}, \label{attachment} \\ 
   \cee{O- + N2 & -> N2O + e}. \label{detachment} 
\end{align}
\end{subequations}
\end{linenomath*}

The electron mobility and the reaction rates for (\ref{n2ion})-(\ref{attachment}) are obtained from the solution of a steady-state Boltzmann equation using BOLSIG+ \citep{Hagelaar2005/PSST}, with the cross sections from \cite{Phelps1985/PhRvA} and \cite{Lawton1978/JChPh}.  For (\ref{detachment}) we use the fit of the data from \cite{Rayment1978/IJMSIP} provided by \cite{Luque2012/NatGe}.  The simulation domain is a cylinder that extends vertically from the ground to an altitude of 100 km and radially to 700 km and we use an uniform cartesian grid with cell sizes $\Delta r = $500 m, $\Delta z = $100 m.

 The charge $Q$ in the thundercloud is modeled as a sphere of radius 0.5 km
located in the central axis of our domain at an altitude $h=\SI{7}{km}$ \citep{Maggio2009/JGRD}.  We simulate the lightning stroke by varying this charge in time as
  \begin{linenomath*}
\begin{equation}
  \od{Q}{t} = I(t) = \frac{Q_{\text{max}}}{\tau_1 - \tau_2}
    \left(\exp(-t/\tau_1)-\exp(-t/\tau_2)\right),
\end{equation}
 \end{linenomath*}
where $Q_{\text{max}}$ is the total charge lowered to the ground and $\tau_1$ and $\tau_2$ are, respectively, the total discharge time and the rise time of the discharge current, which we take to be $\tau_1=\SI{1}{ms}$, $\tau_2=\SI{0.1}{ms}$.  The product $hQ_{\text{max}}$, called charge moment change, determines to a good approximation the electric field imposed on the ionosphere.

We took the air density profile from the US Standard Atmosphere \citep{StandardAtmosphere1976} and our initial electron density follows \cite{Hu2007/JGRD}.  The electrons are balanced by positive ions (about 21\% \ce{O2+} and 79\% \ce{N2+}) to ensure that we start from a neutral charge density.

Weak discharges, with charge moment $hQ_{\text{max}}$ below $\sim$350 C km do not cause significant ionization in the ionosphere.   Hence to study the effect of ionization in the upper atmosphere on the ground-level electric field we consider two relevant cases: a weak discharge where the ionosphere is mostly undisturbed and a strong discharge, where there is significant ionization.  We take $Q_{\text{max}}=\SI{25}{C}$ ($hQ_{\text{max}}=\SI{175}{C.km}$) for the weak discharge and $Q_{\text{max}}=\SI{80}{C}$ ($hQ_{\text{max}}=\SI{560}{C.km}$) for the strong one.

\begin{figure}
\includegraphics[width=0.9\columnwidth]{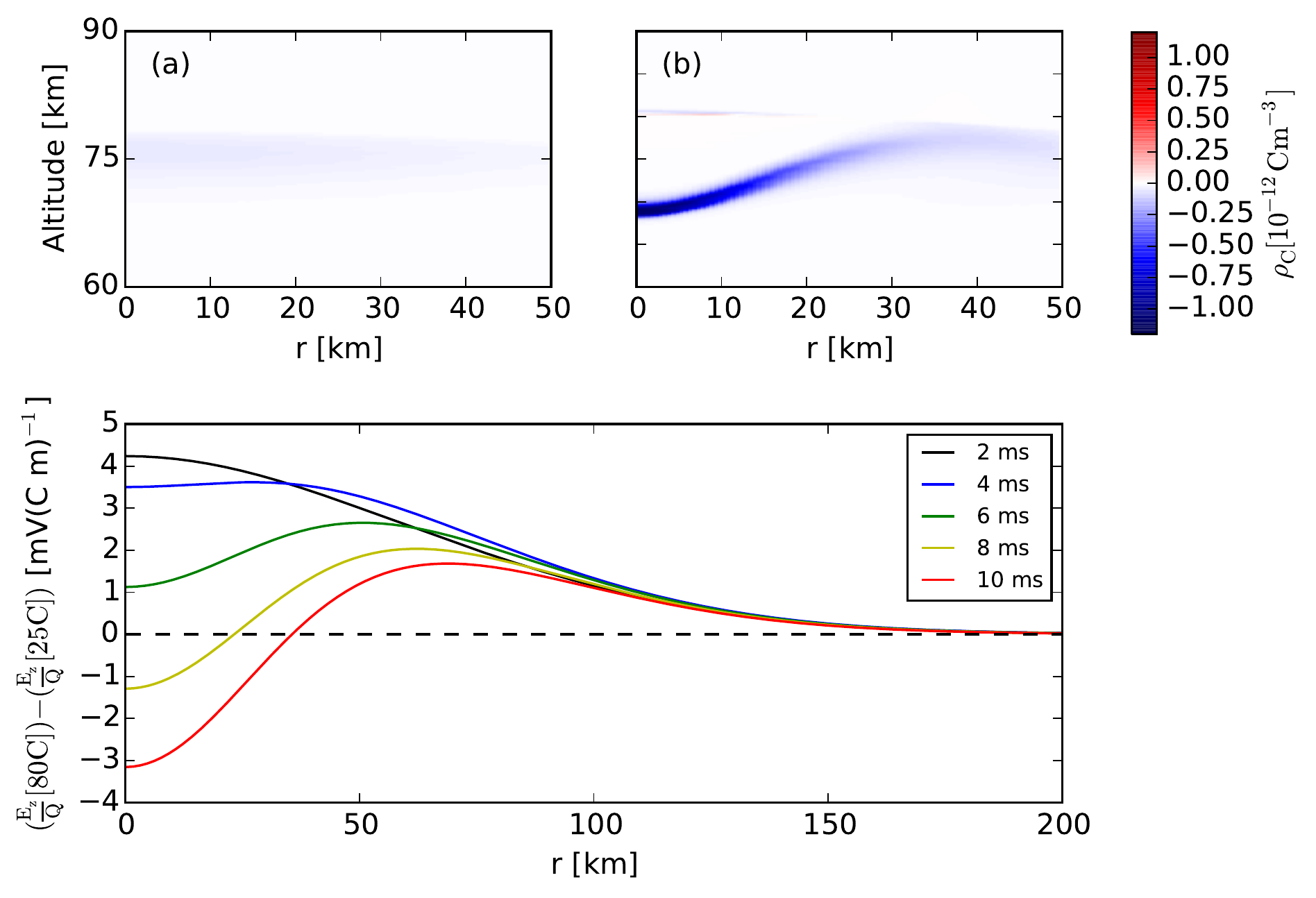}
\caption{\label{fig:charge}
  On the upper panels we can see the space density induced in the lower ionosphere by the thundercloud charge $Q_{\text{max}}$ by (a) a weak discharge of $hQ_{\text{max}}=\SI{175}{C.km}$ and (b) a strong discharge $hQ_{\text{max}}=\SI{560}{C.km}$ causing halo. The total accumulated charge in the lower ionosphere at 10 ms calculated by spatial integration is (a) $Q_\text{i}=-\SI{2.56}{C}$ and (b) $Q_\text{i}=-\SI{8.15}{C}$. On the lower panel we plot the difference between $E$/$Q_{\text{max}}$ at ground level for two different discharges at different times. We can see the halo influence in the first kilometers causing a sign change in the difference.
}
\end{figure}

On the upper panels of figure~\ref{fig:charge} we plot the space charge density induced in the lower ionosphere by each of the two discharges.  For the weak discharge, we see a layer of negative charge around \SI{75}{km} of altitude, which marks the effective altitude of the ionosphere for this case.  The strong discharge creates a bulge in the ionosphere that descends to about \SI{70}{km} within \SI{30}{km} from the axis containing the causative discharge.  Integrating the space charge we find that the accumulated charge in the lower ionosphere is $Q_\text{i}=-\SI{2.56}{C}$ for the weak discharge and $Q_\text{i}=-\SI{8.15}{C}$ for the strong discharge, in good agreement with equation (\ref{Qhalo}) with $L\approx\SI{70}{km}$. 

In figure~\ref{fig:ground-field} we plot the simulated electric field at ground level divided by the total charge lowered to the ground.  We see that the curve is close to that predicted by the analytical expression (\ref{images}) for $L=\SI{70}{km}$.  The collapse of the two simulation profiles in figure~\ref{fig:ground-field} indicates that to a good approximation our results are linear with the driving charge $Q_\text{max}$.  However, there are some factors that break this linearity:
\begin{enumerate}
\item The dependence of the electron mobility with the electric field.  Since electrons are more mobile for low fields, the dielectric relaxation of the ionosphere is somewhat faster if the perturbing field is weaker.  As we argued above, the relaxation of the ionosphere reduces the ground electric field,
so we expect this factor to reduce the ratio $E/Q_\text{max}$ for weak discharges.
\item Changes in the electron density due to the chemical scheme 
  (\ref{reactions}).  A higher electron density accelerates screening and lowers the ionosphere's edge, whereas a lower electron density slows down the screening.  Refering again to our previous arguments, this implies that ionization decreases $E/Q_\text{max}$ whereas attachment increases this ratio.
\end{enumerate}

 On the lower panel of figure~\ref{fig:charge} we compare the ratios $E/Q_\text{max}$ from our model discharges for a range of distances and at several times.  Initially, the effect of the field-dependent mobility dominates and the field is relatively smaller for the weak discharge.  However as ionization lowers down the edge of the ionosphere we see that at short distances the field becomes relatively weaker for the strong discharge, as we argued above.  Far from the discharge the electric fields in the ionosphere are not strong enough for ionization so the effects of attachment and field-dependent mobilities dominate, so $E/Q_\text{max}$ is higher for the strong discharge.  Note however that these nonlinear effects are extremely small, amounting to less than 3\% of the total field. This effect is therefore probably undetectable.

\section{Conclusions}
We have argued that the activity of a halo cannot explain, or at least cannot explain in a straightforward manner, an enhancement in the distant electrostatic field created by a lightning stroke.  We therefore believe that some other explanation is needed for the observations of \cite{Bennett2013/PhRvL}.  At present we do not have a satisfactory alternative but we conclude this letter by listing and discussing some issues that are missing in our models and may provide a path for future investigations.

One such issue is the DC voltage bias between the ground and the ionosphere.  Assuming that the electric field casued by this potential difference is uniform, a decrease of the ionosphere's altitude from say \SI{90}{km} to \SI{70}{km} enhances the fair-weather electric field by a factor $90/70\approx 1.3$, that is, 30\%.  This is a wide upper limit for the increase since the atmospheric conductivity increases exponentially with altitude and the potential difference is located at low altitude.  However, even the 30\% figure looks too small to account for the observed features.

A second issue is the presence of other sources of ionization in the space between the cloud tops and the ionosphere.  As electric fields high enough to cause ionization are also capable of inducing light emissions, the source of ionization that we seek must also be a type of TLE.  Due to their rarity, we can dismiss jets and giant blue jets.  We have sprites and elves as remaining canditates.

Sprites are certainly associated with intense ionization.  To investigate their effects we run simulations where a sprite is modelled as a large, elliptical cloud of ionization above the discharge.  The resulting electrostatic field at ground level is barely distinguishable from the field without a sprite, although slightly smaller.  However, we considered only cases with cylindrical symmetry and therefore did not investigate sprites with a footprint tens of kilometers away from the causative discharge as is often the case \citep{Vadislavsky2009/JASTP}.

The electromagnetic pulse (EMP) emitted by the lightning stroke, visible as an elve as it reaches the lower ionosphere, may also cause significant ionization.  Although usually this ionization increases the conductivity by only a few percent \citep{Marshall2012/JGRA}, in certain extreme cases the increase may be much higher \citep{Gordillo-Vazquez2016/JGRA/temp}.  The highest energy deposition by an EMP interacting with the ionosphere is at a propagation angle of about \ang{45;;} \citep{Luque2014/JGRA}, i.e. at a distance of about \SI{90}{km} from the vertical axis cointaining the causative stroke.  Due to the curvature effect that we discussed above, the ionization caused by the EMP may possibly increase the electrostatic field at ground level at about this distance.  However, we consider EMP-driven ionization an unlikely explanation for the observations of \cite{Bennett2013/PhRvL}: it requires too many intense EMPs and cannot account for any polarity reversal.

Detailed time-resolved and wide-band measurements would greatly illuminate the physics behind the polarity reversal and the violation of the cubic law measured by \cite{Bennett2013/PhRvL}.  This would clarify whether electrostatic fields are a viable alternative for the remote detection and characterization of electric storms.

\section*{Acknowledgement}
This work was supported by the Spanish Ministry of Science and Innovation, MINECO under projects ESP2013-48032-C5-5-R,  FIS2014-61774-EXP and ESP2015-69909-C5-2-R and by the EU through the FEDER program. FJPI acknowledges a MINECO predoctoral contract, code BES-2014-069567. AL acknowledges support by a Ram{\'o}n y Cajal contract, code RYC-2011-07801. All data used in this paper are directly available after a request is made to authors FJPI (fjpi@iaa.es), AL (aluque@iaa.es) or FJGV (vazquez@iaa.es).

\newcommand{\jcp}{J. Chem. Phys. } 
\newcommand{\ssr}{Space Sci. Rev.} 
\newcommand{\planss}{Plan. Spac. Sci.} 
\newcommand{\pre}{Phys. Rev. E} 
\newcommand{\nat}{Nature} 
\newcommand{\icarus}{Icarus} 
\newcommand{\ndash}{-} 
\newcommand{\jgr}{J. Geophys. Res.}

%\end{article}
\end{document}